\documentclass[a4paper,11pt]{article}
\pdfoutput=1
\usepackage{jcappub}

\newcommand{\be}{\begin{equation}}
\newcommand{\ee}{\end{equation}}
\newcommand{\res}{\text{res}}
\newcommand{\textmax}{\text{max}}
\newcommand{\order}{\mathcal{O}}
\title{Constraining ALP-photon coupling using galaxy clusters}

\author[a]{Martin Schlederer,}
\author[a]{G\"unter Sigl}

\affiliation[a]{II. Institut f\"ur theoretische Physik, Universit\"at Hamburg
\\
Luruper Chaussee 149, 22761 Hamburg, Germany}

\emailAdd{martin.schlederer@desy.de}
\emailAdd{guenter.sigl@desy.de}

\abstract{
We study photon-ALP conversion by resonance effects in the magnetized plasma of galaxy clusters and compare the predicted distortion of the cosmic microwave background spectrum in the direction of such objects to measurements of the thermal Sunyaev-Zeldovich effect.
Using galaxy cluster models based on current knowledge, we obtain upper limits on the photon-ALP coupling constant $g$ of $\lesssim \mathcal{O}(10^{-11}$ GeV$^{-1}$).
The constraints apply to the mass range of $2\cdot 10^{-14}$ eV $ \lesssim m_\text{ALP} \lesssim 3\cdot 10^{-12}$ eV in which resonant photon-ALP conversions can occur.
These limits are slightly stronger than current limits, and furthermore provide an independent constraint.
We find that a next generation PRISM-like experiment would allow limits down to $g \approx \mathcal{O} (10^{-14}$ GeV$^{-1})$, two orders of magnitude stronger than the currently strongest limits in this mass range.}

\begin{document}
\maketitle
\flushbottom

%%%%%%%%%%%%%%%%%%%%%%%%%%%%%%%%%%%%%%%%%%%%%%%%%%%%%%%%%%%%%%%%%%%%%%%%%%%%%%%%%%%%%%%%%%%%%%%%%%%%%%%%%%%
\section{Introduction}

While a lot of effort is put into the search for new physics at large accelerators like the LHC at CERN, another approach is to instead search for new physics at very low energy scales and small couplings.
In this context, the axion is one of the best known candidates: it was introduced in 1977 by Peccei and Quinn to solve the strong CP problem \cite{pecceiQuinn}, but despite all effort, it has not been found yet, and both its mass and its coupling to photons are still unknown.
In addition to the axion, several extensions of the standard model predict similar particles, so called ``axion-like particles'' (ALPs) (see, e.g., \cite{ringwald:2010} for a review).
But while the axion must satisfy a certain relation between its mass and its coupling to photons in order to solve the strong CP problem,
in general there is no relation between mass and coupling constants for ALPs.
These ALPs have been suggested to explain several physical phenomena \cite{ringwald:2014}, such as the anomalous gamma-ray transparency \cite{DeAngelis:2007, Horns:2012}, the soft X-ray excess from the Coma cluster \cite{Conlon:2014}, or dark matter \cite{Preskill:1983, Abbott:1983, Dine:1983, Arias:2012}.
Interestingly, both the anomalous gamma-ray transparency and the soft X-ray excess from the Coma cluster may be explained with ALPs in a similar parameter region: the gamma-ray transparency can be resolved with a photon-ALP coupling constant $g \gtrsim 10^{-(10-11)}$ GeV$^{-1}$ and an ALP mass $m_\phi \lesssim 10^{-7}$ eV \cite{Meyer:2013}, while the soft X-ray excess can be explained with $g \gtrsim 10^{-13}$ GeV$^{-1}$ and $m_\phi \lesssim 10^{-12}$ eV \cite{Angus:2014}.
\\
In this work, we propose a method to improve current limits in the mass region of $10^{-14}$ eV $\lesssim m_\phi \lesssim 10^{-12}$ eV, therefore reducing the parameter space available for explaining the soft X-ray excess and the gamma-ray transparency with ALPs:
for a suitable ALP mass, CMB photons crossing a galaxy cluster can undergo resonant photon-ALP conversion inside the cluster's magnetic field, therefore distorting the black-body spectrum of the CMB.
Galaxy clusters are one of the few large scale astrophysical objects with known magnetic fields which allows to derive constraints not only on the combination $gB$ from distortions of the CMB \cite{resPhotonAxion}, but also on $g$ itself.
In particular, we use observations of the thermal Sunyaev-Zeldovich effect by OVRA, WMAP, MITO, and the \textit{Planck} satellite
and use these to obtain limits on the coupling between ALPs and photons.
While current data leads to limits only slightly better than the ones obtained from SN1987A, a future PRISM-like experiment might significantly improve current limits.
The currently strongest limits in this mass-region are derived from the absence of a gamma-ray flash at the time of the SN1987A and limits the coupling to photons to $g\lesssim 5.3 \cdot 10^{-12}$ GeV$^{-1}$ \cite{sn1987:3}.
\\
This paper is structured as follows: in section 2 we introduce the framework of resonant photon-ALP conversion inside galaxy clusters
and describe the cluster models used. 
The resulting constraints are presented in section 3.
This section also contains an estimate of the expected sensitivity of a PRISM-like experiment.
In section 4 we discuss the results and how they may be further strengthened.
In the appendix we present in detail how the multiple level crossing has been calculated.
\\
Throughout the paper, we set $c = \hbar = k_B = 1$. We denote spatial vectors with bold face symbols.

%%%%%%%%%%%%%%%%%%%%%%%%%%%%%%%%%%%%%%%%%%%%%%%%%%%%%%%%%%%%%%%%%%%%%%%%%%%%%%%%%%%%%%%%%%%%%%%%%%%%%%%%%%%
\section{Framework of photon-ALP oscillations}
In this chapter, we derive an expression for the conversion probability from photons to ALPs and compare the corresponding temperature change with observations of the thermal Sunyaev-Zeldovich effect. 
Using suitable models for the profile of galaxy clusters, we then determine upper limits for the coupling constant between photons and ALPs.

%%%%%%%%%%%%%%%%%%%%%%%%%%%%%%%%%%%%%%%%%%%%%%%%%%%%%%%%%%%%%%%%%%%%%%%%%%%%%%%%%%%%%%%%%%%%%%%%%%%%%%%%%%%
\subsection{Resonant photon-ALP conversion and its effect on the CMB temperature}
For the deduction of the conversion probability we closely follow \cite{resPhotonAxion}.
Axion-like particles (ALPs), in this work denoted by $\phi$, are pseudoscalar bosonic particles, that couple to photons through the interaction Lagrangian \cite{photonALPLagrangian}
\begin{equation}
\mathcal{L} = - \frac{1}{2}g F_{\mu \nu} \tilde{F}^{\mu \nu} \phi= g \mathbf{B} \cdot \mathbf{E} \phi,
\end{equation}
where $\tilde{F}^{\mu \nu} = \epsilon^{\mu \nu \alpha \beta} F_{\alpha \beta}/2$ is the dual of the electromagnetic field strength tensor, $\mathbf{B}$ and $\mathbf{E}$ are the magnetic and electric field, respectively, and $g$ denotes the axion-photon coupling constant.
\\
In an external magnetic field, this interaction Lagrangian is well known to produce effective mass-mixing between photons and ALPs. The new propagation eigenstates are then rotated with respect to the interaction eigenstates by an angle $\theta$ given by \cite{raffelt:1988}
\begin{equation}
\label{eq:thetaNoPlasma}
\sin 2\theta = \frac{2gB\omega}{\sqrt{m_{\phi}^4 + (2gB\omega)^2}}, \qquad
\cos 2\theta = \frac{m_{\phi}^2}{\sqrt{m_{\phi}^4 + (2gB\omega)^2}}.
\end{equation}
Here, $\omega$ denotes the energy of the photon, $B$ is the component of the magnetic field perpendicular to the propagation direction of the photons, $m_{\phi}$ is the ALP mass and $g$ is the coupling constant as before.
From here on, we will refer to $\theta$ as magnetic mixing angle.
\\
Using typical values for the parameters in our study, the relevant dimensionless parameter in these expressions reads
\begin{equation}
\frac{2gB\omega}{m^2_\phi} \simeq 1.38 \cdot 10^{-6} \frac{g}{10^{-12} \mbox{ GeV}^{-1}} \frac{B}{\mu \mbox{G}} \frac{\omega}{10^{-4} \mbox{ eV}} \left( \frac{10^{-13} \mbox{ eV}}{m_\phi} \right)^2.
\end{equation}
For the parameter ranges considered here this will always be much smaller than unity.
The misalignment between interaction eigenstates and propagation eigenstates will produce photon-ALP oscillations with a wavenumber given by \cite{raffelt:1988}
\begin{equation}
\label{eq:k}
k = \frac{\sqrt{m_\phi^4 + (2gB \omega)^2}}{2 \omega}.
\end{equation}
Inside a plasma, photon-ALP mixing will be modified: the refractive properties of the plasma lead to a non-trivial dispersion relation, which can be parametrized by an effective photon mass $m_\gamma$.
In this case, for a given magnetic field the effective mixing angle in the plasma $\tilde{\theta}$ is related to the mixing angle $\theta$ at vanishing charge density by \cite{raffelt:1996}
\begin{equation}
\label{eq:angleInPlasma}
\sin 2\tilde{\theta} = \frac{\sin 2 \theta}{[\sin^2 2 \theta + (\cos 2 \theta - \xi)^2 ]^{1/2}},
\end{equation}
\begin{equation}
\cos 2\tilde{\theta} = \frac{\cos 2 \theta - \xi}{[\sin^2 2 \theta + (\cos 2 \theta - \xi)^2 ]^{1/2}},
\end{equation}
where $\xi$ is defined as
\begin{equation}
\xi \equiv \cos 2 \theta \left( \frac{m_\gamma}{m_\phi} \right)^2.
\end{equation}
If in some region in space the resonance condition
\begin{equation}
\label{eq:resonanceCondition}
m_\gamma = m_\phi
\end{equation}
is satisfied, one has $\tilde{\theta} \rightarrow \pi/4$ and resonant photon-ALP conversion is possible.
\\
In this study, we will consider such resonant photon-ALP conversion occurring in galaxy clusters, where one has both
external magnetic fields and a nonzero free electron density.
\\
Due to the high ionization fraction in the intra-cluster medium, contributions from the scattering off neutral atoms can be neglected and the effective photon mass is given by the plasma frequency \cite{born:1980}
\begin{equation}
\label{eq:m_gamma}
m_\gamma^2 \simeq \omega_P^2 = \frac{4 \pi \alpha}{m_e} n_e,
\end{equation}
where $\alpha$ is the fine structure constant, $m_e$ is the electron mass and $n_e$ is the free electron density.
Using this relation, one can rewrite the resonance condition (\ref{eq:resonanceCondition}) to
\begin{equation}
m_\phi = 3.72 \left( \frac{n_e}{\mbox{m}^{-3}} \right)^{1/2} \cdot 10^{-14}~\mbox{eV}.
\end{equation}
As the free electron density cannot reach arbitrary values inside a galaxy cluster, the resonance condition can only be satisfied 
for a certain range of ALP masses $m_\phi$.
Assuming for example a minimal density of $0.3$ m$^{-3}$ and a maximal density of $10^{4}$ m$^{-3}$,
resonant photon-ALP conversion will only occur for $2 \cdot 10^{-14}$ eV $\lesssim m_\phi \lesssim 3.7 \cdot 10^{-12}$ eV.
The first value is the average baryon density in the universe at redshift zero and the second value is a typical density in the core of a galaxy cluster.
\\
The distance between the photon production and the resonance as well as the distance between the resonance and the detection is much larger than the oscillation length causing an incoherent superposition of the oscillation patterns.
In this case, the transition probability is given by \cite{parke:1986}
\begin{equation}
\label{eq:generalConversionFormula}
P_{\gamma \rightarrow \phi} \simeq \frac{1}{2} + \left( p - \frac{1}{2} \right) \cos 2\tilde{\theta}_0 \cos 2 \tilde{\theta}_D,
\end{equation}
where $\tilde{\theta}_0$ is the effective mixing angle at production, $\tilde{\theta}_D$ is the effective mixing angle at detection and $p$ is the level crossing probability. Therefore, a transition from a medium dominated to the vacuum state corresponds to
$\cos2\tilde\theta_0\simeq-1$ and $\cos2\tilde\theta_D\simeq\cos2\theta$ or \textit{vice versa} and for $\theta\ll1$ one obtains
a conversion probability $P_{\gamma \rightarrow \phi}$ close to unity for $p\ll1$, corresponding to an adiabatic transition.
\\
As it was argued in \cite{resPhotonAxion}, the high plasma density at the time of the creation of the CMB photons leads to a value of $\tilde{\theta}_i$ close to $\pi/2$.
Using typical values of the free electron density in the solar system, one can see that also $\tilde{\theta}_D$ is very close to $\pi/2$.
A more detailed discussion of this point can be found in the appendix.
\\
For our range of ALP masses, there is not only one, but several resonances:
the first resonance occurs when the free electron density decreases due to the cosmic expansion.
Inside the transversed cluster, there are several resonances: additionally to the two resonances due to the increasing (decreasing) electron density when entering (leaving) the cluster, density fluctuations enable even more resonances.
Finally, there are also resonances when the photons enter the Milky Way.
\\
The level crossing probability $p_i$ for a single resonance $i$ is given by the Landau-Zener expression \cite{kuo:1989}
\begin{equation}
\label{eq:LandauZener}
p_i \simeq \exp(-2\pi R k \sin^2\theta_{\text{res}}),
\end{equation}
where $k$ is the oscillation wavenumber given in eq. (\ref{eq:k}), $\theta_{\text{res}}$ is the magnetic mixing angle given by eq. (\ref{eq:thetaNoPlasma}) at this resonance and $R$ is the scale parameter defined as
\begin{equation}
\label{formula:scaleParameter}
R = \left| \frac{d \ln m^2_\gamma(t)}{dt} \right|^{-1}_{t=t_{\text{res}}}.
\end{equation}
The level crossing probability $p_i$ takes into account the deviation from adiabaticity of the photon-ALP conversion in the resonance region. One has $p_i \simeq 0$ for a completely adiabatic transition and $p_i = 1$ for an extremely non-adiabatic one.
\\
The Landau-Zener expression (\ref{eq:LandauZener}) only holds for the case when the free electron density varies linearly during the resonance.
For $\theta_{\rm res}$ the resonance half-width is, according to equation (\ref{eq:angleInPlasma}), $\delta \xi \simeq \sin 2 \theta_{\text{res}}$, corresponding to a resonance width in the density scale of
\be
\Delta n_e^R \simeq n_e^R \sin(2\theta_\res) \simeq n_e^R \frac{2gB\omega}{m_\phi^2}
\simeq 10^{-6} n_e^R,
\ee
where $n_e^R$ is the resonance density defined by eq. (\ref{eq:resonanceCondition}).
Due to this extremely narrow resonance region, the approximation of a linear density change is very well fulfilled.
\\
Expanding the sine in eq. (\ref{eq:LandauZener}) and approximating $k \simeq m_\phi^2/(2\omega)$, the exponent becomes
\be
2\pi Rk \sin^2(\theta_\res) \simeq \frac{g^2B^2_\res R \omega \pi}{m_\phi^2}
\simeq \order(10^{-6}),
\ee
for typical values used in this work.
As derived in Appendix A, for such small exponents, the total level crossing probability becomes
\be
p = \prod_i p_i \simeq 1 - \sum_i \frac{g^2 B^2_i R_i \omega \pi}{m_\phi^2},
\ee
where the sum is over all resonances $i$.
Together with the obtained expressions for $\cos 2 \tilde{\theta}_0$, and $\cos 2 \theta_D$, the conversion probability then is
\be
\label{formula:convProb}
P_{\gamma \rightarrow \phi} \simeq \sum_i \frac{g^2 B^2_i R_i \omega \pi}{m_\phi^2} = \frac{g^2\omega \pi}{m_\phi^2} \sum_i B_i^2 R_i.
\ee
In the present study we will neglect the first resonance because of the unknown magnetic field, and the  resonances in the Milky Way because of the small scale parameter $R$.
These approximations are conservative, since they tend to underestimate the actual conversion probability, as it will be discussed in Appendix A.
\\
The conversion of photons into ALPs will always reduce the number of photons in the beam, causing an apparent temperature decrease.
The intensity for a given photon energy and temperature $I(\omega, T)$ can be related with the apparent temperature change over
\begin{equation}
\label{formula:deltaT}
I' = \frac{dI(\omega, T)}{dT} \simeq \frac{\Delta I}{\Delta T} ~\rightarrow~ \Delta T(\omega) \simeq \frac{\Delta I(\omega, T)}{I'(\omega, T)}.
\end{equation}
The intensity of the photon beam after crossing the galaxy cluster is reduced by a factor of $1-P_{\gamma \rightarrow \phi}$, giving $\Delta I(\omega, T) = -P_{\gamma \rightarrow \phi} I_0(\omega, T)$.
With $I(\omega, T) \propto \left[ \exp(\omega / T_\text{CMB} ) -1 \right]^{-1}$
one arrives at
\begin{equation}
\label{formula:DeltaTFromConversion}
\Delta T_{\gamma \rightarrow \phi}(\omega) = - P_{\gamma \rightarrow \phi} \frac{T^2_{CMB}}{\omega} \left[ 1 - \exp \left( -\frac{\omega}{T_{\text{CMB}}} \right) \right]
\end{equation}
for the apparent temperature change in dependence of the photon energy due to resonant photon-ALP conversion.

%%%%%%%%%%%%%%%%%%%%%%%%%%%%%%%%%%%%%%%%%%%%%%%%%%%%%%%%%%%%%%%%%%%%%%%%%%%%%%%%%%%%%%%%%%%%%%%%%%%%%%%%%%%
\subsection{Comparing the conversion probability with the tSZ-parameter}
\label{section:comparingTSZandPAC}

From the \textit{Planck} 2015 data \cite{planck2015SZ}, we have information about the temperature differences in the directions of galaxy clusters. 
These temperature differences depend on the frequency observed and are related to the (by definition frequency-independent) thermal Sunyaev-Zeldovich (tSZ) Compton parameter $y$ by the expression \cite{sunyaev:1972}
\begin{equation}
\label{formula:tSZ}
\Delta T_{\text{tSZ}}(\omega) = f(\omega)~ T_{\text{CMB}} ~ y,
\end{equation}
with $f(\omega) = (\omega/T_{\text{CMB}})\cdot \coth [ \omega/(2 T_{\text{CMB}}) ] - 4$.
The function $f(\omega)$ is negative for values of $\omega$ smaller than $3.83\,T_{\text{CMB}}$, otherwise the function is positive.
%%%%%%%%%%%%%%%%%%%%%%%%%%%%%%%%%%%%%%%%%%%%%%%%%%%%%%%%%%%%%%%%%%%%%%%%%%%%%%%%%%%%%%%%%%%%%%%%%%%%%%%%%%
%%%%%%%%%%%%%%%%%%%%%%%%%%%%%%%%%%%%%    FIGURE    %%%%%%%%%%%%%%%%%%%%%%%%%%%%%%%%%%%%%%%%%%%%%%%%%%%%%%%
%%%%%%%%%%%%%%%%%%%%%%%%%%%%%%%%%%%%%%%%%%%%%%%%%%%%%%%%%%%%%%%%%%%%%%%%%%%%%%%%%%%%%%%%%%%%%%%%%%%%%%%%%%
\begin{figure}[t]
\centering
{
\includegraphics[width=10cm]{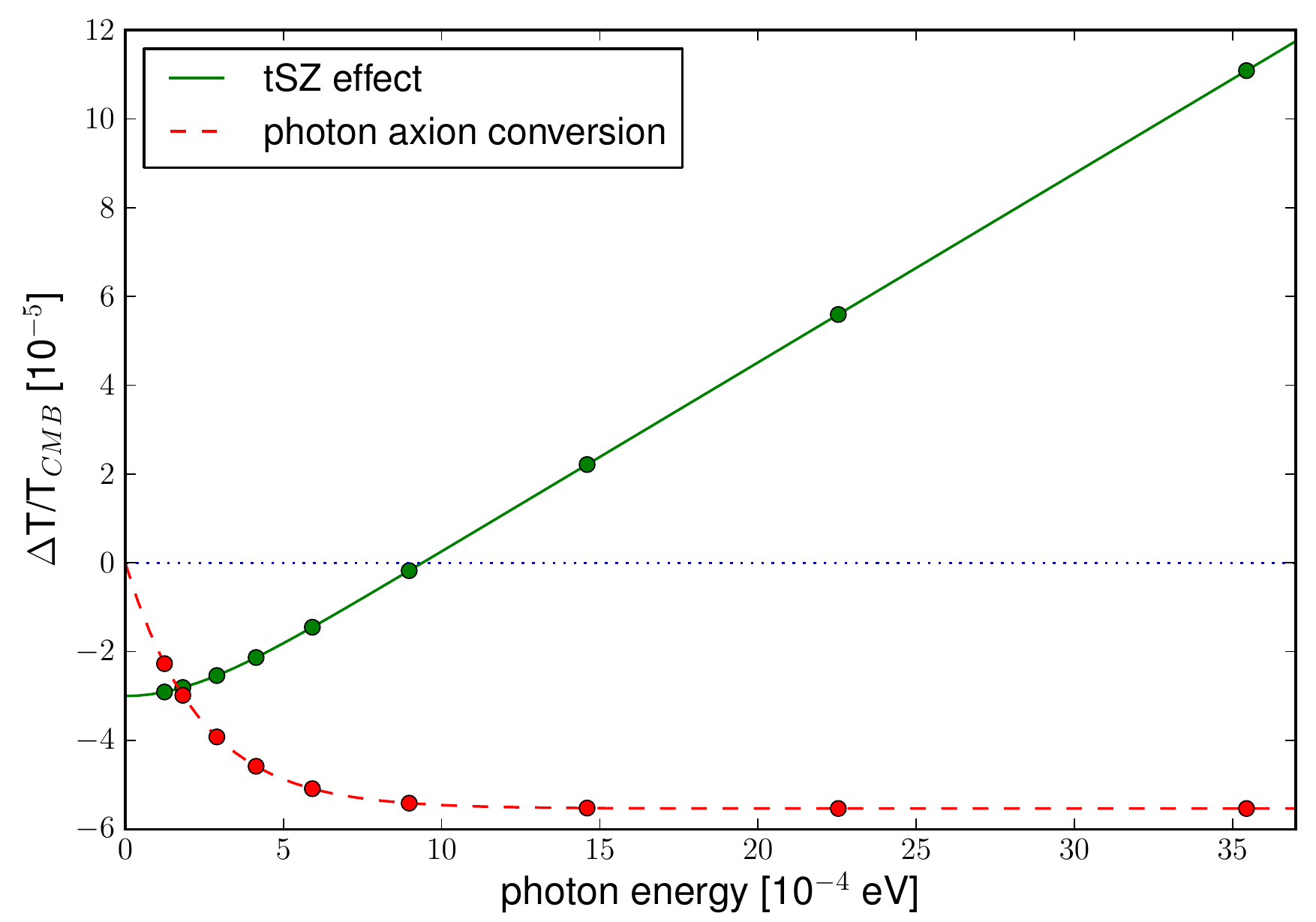}
}
\caption{Comparison of the spectral profile of the tSZ effect (with y = $10^{-5}$) and the photon-axion conversion.
The $x-$axis shows the photon energy, the $y-$axis shows the relative distortion of the effective CMB temperature,
$\Delta T/T_{\text{CMB}}$ in units of $10^{-5}$.
For this plot, the parameters $g = 5\cdot 10^{-13}$ GeV$^{-1}$, $B$ = 2 $\mu$G, $R$ = 0.5 Mpc, $m_{\phi} = 10^{-13}$ eV were used.
Different values for these parameters change the normalization of the effect but not its dependence on the photon frequency.
The filled circles refer to the centers of the frequency bands of the \textit{Planck}-mission.}
\label{tSZandConversion}
\end{figure}
%%%%%%%%%%%%%%%%%%%%%%%%%%%%%%%%%%%%%%%%%%%%%%%%%%%%%%%%%%%%%%%%%%%%%%%%%%%%%%%%%%%%%%%%%%%%%%%%%%%%%%%%
In figure (\ref{tSZandConversion}) the relative temperature change due to the thermal SZ effect as well as due to photon-ALP conversion is shown.
As mentioned above, the relative temperature change due to the photon-ALP conversion is always negative, because the effect always removes photons from the beam.
The tSZ-effect, in contrast, creates a negative temperature change for small energies ($\omega < 3.83~ T_{\text{CMB}}$) and a positive temperature change at higher photon energies.
\\
For the Coma cluster, detailed measurements of its thermal SZ-effect for photon energies in the range of $10^{-4}$ eV $ \lesssim \omega \lesssim 1.1 \cdot 10^{-3}$ eV exist \cite{battistelli:2003}.
These values are presented in table (\ref{table:tSZdata}).
\begin{table}[t]
\begin{center}
\begin{tabular}
 {|l|c|r|} \hline Experiment & $\omega$ [10$^{-4}$ eV] & $\Delta$ T$_{tSZ}$ [$\mu$K]\\
\hline \hline
OVRA & 1.32 & -520 $\pm$ 83 \\ \hline
WMAP & 2.51 & -240 $\pm$ 180 \\ \hline
WMAP & 3.87 & -340 $\pm$ 180 \\ \hline
MITO & 5.91 & -184 $\pm$ 39 \\ \hline
MITO & 8.85 & -32 $\pm$ 79 \\ \hline
MITO & 11.25 & 172 $\pm$ 36 \\ \hline
\end{tabular}
\end{center}
\caption{Measurements of the thermal SZ-effect in the Coma cluster.}
\label{table:tSZdata}
\end{table}
In this case, one can build the reduced $\chi^2$-function
\begin{equation}
\label{formula:chiSqTest}
 \chi^2(y, g) = \frac{1}{N-2} \sum_{i}^{N} \left[ \frac{\Delta T_i^{\text{exp}} - \Delta T_i^{\text{theo}}(y, g)}{\sigma_i^{\text{exp}}} \right]^2,
\end{equation}
where $\Delta T_i^{\text{exp}}$ are the observed temperature changes at photon energy $\omega_i$, $\sigma_i$ are their standard errors and
\begin{equation}
 \Delta T_i^{\text{theo}}(y, g) = \Delta T_{\gamma \rightarrow \phi}(\omega_i, g) + \Delta T_{\text{tSZ}}(\omega_i, y),
\end{equation}
is the prediction by the theory, with $\Delta T_{\gamma \rightarrow \phi}$ and $\Delta T_{\text{tSZ}}$ given by eq. (\ref{formula:DeltaTFromConversion}) and eq. (\ref{formula:tSZ}), respectively.
For each ALP mass $m_{\phi}$, we calculate the $\chi^2$-function in the ($y, g$)-parameter space and determine the limit on $g$ as the largest value of it still inside the respective confidence interval.
\\
Often, e.g. in the case of the \textit{Planck}-mission \cite{planck2015SZ}, only the $y$-parameters of galaxy clusters are easily available, but not the temperature changes due to the thermal SZ-effect at different frequencies.
In such cases, a simple and conservative bound can be obtained when assuming that the magnitude of the temperature change due to the photon-ALP conversion must be smaller than the magnitude of the temperature change due to the tSZ-effect:
\begin{equation}
\label{formula:temperatureComparison}
| \Delta T_{\gamma \rightarrow \phi} | \lesssim | \Delta T_{\text{tSZ}} |
\end{equation}
Using equations (\ref{formula:convProb}), (\ref{formula:DeltaTFromConversion}), (\ref{formula:tSZ}), and solving for the coupling constant $g$, one arrives at
\begin{equation}
\label{formula:gBound}
g \lesssim m_\phi \left( \frac{y}{\sum_i B_i^2 R_i} \right)^{1/2} h(\omega), 
\end{equation}
where we defined
\be
\label{formula:h}
h(\omega) \equiv \left| \frac{4-x\cdot \coth(x/2)}{T_\text{CMB} \pi [1 - \exp(-x)]} \right| ^{1/2}, ~~ x \equiv \frac{\omega}{T_{\text{CMB}} }.
\ee
Note that this approach does not take any non-resonant conversion into account, and therefore conservatively overestimates $g$.

%%%%%%%%%%%%%%%%%%%%%%%%%%%%%%%%%%%%%%%%%%%%%%%%%%%%%%%%%%%%%%%%%%%%%%%%%%%%%%%%%%%%%%%%%%%%%%%%%%%%%%%%%%%
\subsection{Galaxy cluster model and multiple resonances}
Both the magnetic field strength as well the scale parameter at resonance depend on the free electron density inside the considered galaxy cluster.
On top of the smooth, large-scale electron density profile, galaxy clusters exhibit smaller turbulent contributions in the electron density.
For example, in the Coma cluster, density fluctuations of $\sim$5\% on scales of $\sim$30 kpc and $\sim$(7-10)\% on scales of $\sim$500 kpc have been observed \cite{churazov:2012}.
As estimated before, in terms of the electron density variation the resonance is extremely narrow,
$\Delta n_e/n_\res \approx 10^{-6}$, such that small turbulent contributions can already cause several distinct resonances.
We therefore have to evaluate the expression $\sum_i B_i^2 R_i$ from eq. (\ref{formula:convProb}), where the index $i$ labels the individual resonances.
\\
We assume that the magnetic field strength follows the free electron density, such that
\begin{equation}
\label{eq:magneticFieldWithDensity}
\left< | \textbf{B}(r) | \right> \simeq B_{\text{max}} [ n_e(r) / n_{\text{max}} ]^\eta,
\end{equation}
where $\eta \lesssim \order(1)$ and $B_\text{max}$ will be specified later.
Due to the narrowness of the resonance region, one can approximate the magnetic field as being constant during the resonance.
Furthermore, the magnetic field during the resonance is completely determined by the resonance density $n_\res$, such that all resonances will experience the same field strength.
Averaging over several resonances, one therefore obtains
\be
\sum_i B_i^2 R_i \simeq \frac{2}{3} B_\res^2\sum_i R_i,
\ee
where the factor of 2/3 accounts for the fact that only the two transverse components of the magnetic field enter the conversion probability (\ref{formula:convProb}).
\\
To investigate the consequences of turbulent density contributions, we will consider an electron density with a dominant smooth component $n_\text{s}$ and a spectrum of modulations with wavenumber $k$, amplitude $\delta(k)$, and phase $\phi_k$, such that
\be
n_e(r) = n_\text{s}(r) \cdot \prod_k [1+\delta(k) \sin(kr+\phi_k) ].
\ee
We will assume some upper cutoff $k_\textmax$ induced by viscous damping and $\delta(k) \ll 1$ as indicated by observation.
As $\Delta n_\res / n_\res \ll 1$, one can neglect ``incomplete'' resonances (where the density has an extremum), but always assume the electron density to vary linearly during the resonance.
The scale parameter $R_i$, defined by eq. (\ref{formula:scaleParameter}), of an individual resonance $i$ at radius $r_i$ then is
\be
\label{eq:scaleParameterSpectrum}
R_i = \Big| \frac{d \ln[n_\text{s}(r)]}{dr} + \sum_k \frac{\delta(k) k \cos(kr+\phi_k)}{1+ \delta(k) \sin(kr + \phi_k)} \Big|_{r = r_i}^{-1}.
\ee
Using $\delta($500 kpc) $\sim(7-10)\%$, $\delta($30 kpc) $\sim5\%$ from the Coma cluster, and assuming a power law $\delta(k) \propto k^{-\xi}$ provides $\xi\simeq 0.12 ... 0.25$, implying $\delta(k) k \propto k^{0.75...0.88}$.
The scale parameter will therefore be dominated by the contribution from the largest wavenumber not affected by damping,
while, due to the random phases $\phi_k$, the contributions from larger scales approximately average out and can be neglected.
With $\delta(k) \ll 1$, and denoting the dominating wavenumber as $k_\text{dom}$, one thus has
\be
\label{eq:RdominantWavenumber}
R_i \simeq \Big| \frac{d \ln[n_s(r)]}{dr} + \delta(k_\text{dom}) k_\text{dom} \cos(k_\text{dom}r_i + \phi_k) \Big|^{-1}.
\ee
The spatial width $\Delta r$ containing all the resonances can be estimated by the width, within which the smooth profile varies by a factor of ($1 \pm \delta_\textmax$), where $\delta_\textmax$ is the largest modulation amplitude.
Explicitly, $\Delta r \simeq 2 \delta_\textmax n_\res / |n_\text{s}'|$, where $n_\text{s}' = dn_\text{s}(r)/dr|_{\text{res}}$ is the derivative of the smooth profile, evaluated at resonance.
As $\delta(k_\text{dom}) \leq \delta(k_\textmax)$, one can conservatively set
\be
\label{formula:DeltaR}
\Delta r \simeq 2 \delta(k_\text{dom}) n_\text{res} / |n_\text{s}'|,
\ee
neglecting a factor of order unity.
Including additional modes would enable resonances within an even larger region, making this statement only more conservative.
\\
The number of resonances can then be estimated by
\be
\label{eq:numberofresonances}
N \simeq 2 \frac{k_\text{dom} \Delta r}{2 \pi} \simeq \frac{2\delta(k_\text{dom}) k_\text{dom} n_\res}{\pi |n_\text{s}'|},
\ee
where the factor of two in the first equality arises because there are two resonances per complete period.
Note that there must always be at least one resonance (if $n_\res < n_\text{max}$) for radially incoming (outgoing) photons due to the increasing (decreasing) electron density.
\\
As $N$ has to be a natural number, this formula is only a good approximation for $N \gg 1$, while, for $N \approx 1$, we expect an uncertainty of order unity.
The assumption $N \gg 1$ implies certain conditions on the density fluctuations:
density fluctuations of $\sim 5\%$ on scales of 30 kpc have been detected in the Coma cluster, providing $N \gtrsim 2$, while projection effects preclude strong limits for smaller scales \cite{churazov:2012}.
Rotation measures, however, indicate that the Coma cluster contains magnetic fields with coherence lengths down to $\sim2$ kpc \cite{bonafede:2010}.
Due to turbulence, one expects density fluctuations on similar length scales.
Again using $\delta(k) \propto k^{-\xi}$ with $\xi\simeq 0.12 ... 0.25$, one obtains $\delta($2 kpc) $\simeq(2.5...3.6)\%$, while $N \gg 1$ implies $\delta$(2 kpc) $\gg 0.2\%$.
In the last inequality, we assumed the $\beta$-model introduced below as a smooth profile and the parameters of the Coma cluster.
In the same cluster, and $\delta($2 kpc) $\sim 3\%$, one obtains $N \gtrsim 15$.
For such a high number of resonances, the second term in (\ref{eq:RdominantWavenumber}) dominates, and one can approximate the sum over all resonances by averaging the trigonometric functions, leading to
\be
\sum_i R_i \simeq \sum_i \Big| \delta(k_\text{dom}) k_\text{dom} \cos(k_\text{dom}r_i + \phi_k) \Big|^{-1} \simeq N\frac{\pi}{2\delta(k_\text{dom}) k_\text{dom}} \simeq \frac{n_\res}{|n_\text{s}'|} = R_\text{s},
\ee
where $\langle | \cos(x) | \rangle = 2/\pi$ and eq. (\ref{eq:numberofresonances}) for $N$ has been used.
$R_\text{s}$ is the scale parameter one obtains from the smooth profile without density modulation.
\\
If $N$ is $\order(1)$, the scale parameter (\ref{eq:scaleParameterSpectrum}) becomes $R_i \approx R_\text{s} / N$.
One could then still approximate $\sum_i R_i \simeq R_\text{s}$, inducing a relative error of $\lesssim \mathcal{O}(1)$.
\\
As we have seen, the exact dominating wavenumber $k_\text{dom}$ and corresponding amplitude $\delta(k_\text{dom})$ do not influence the transition probability as long as $N \gg 1$.
We will therefore not specify them in any more detail and approximate
\be
\label{eq:finalConvProb}
\sum_i B_i^2 R_i \simeq \frac{2}{3} B_\res^2 R_\text{s}.
\ee
Independent of the multiple resonances due to the turbulent structure, there is one region of resonance when the photons enter the galaxy cluster, and one region of resonances when the photons leave the cluster.
Thus, an additional factor of 2 has to be included in the conversion probability.
In total, one therefore obtains
\be
\label{eq:convProb2}
P_{\gamma \rightarrow \phi} \simeq \frac{4g^2 B_\res^2 R_\text{s} \omega \pi}{3m_\phi^2}.
\ee
For numerical calculations, we will consider a $\beta$-profile as the dominant smooth profile
\be
\label{formula:betaProfile}
n_\text{s}(r) = \max \left( n_0,~ n_{\text{max}} [ 1 + (r / r_c)^2]^{-3 \beta / 2} \right),
\ee
where $\beta$ is $\mathcal{O}$(1), $n_{\text{max}}$ is the free electron density in the cluster center, and $r_c$ is the core radius of the cluster.
We include the average cosmological electron density $n_0$ as a lower boundary for the electron density, and therefore, due to eq. (\ref{eq:resonanceCondition}), a lower cutoff for the ALP-masses $m_\phi$ able to undergo resonance.
\\
We furthermore will focus on two clusters: the Coma cluster and the Hydra A cluster.
In \cite{bonafede:2010}, rotation measure images have been used to determine the Coma clusters magnetic field strength as well as the parameter $\eta$, defined by eq. (\ref{eq:magneticFieldWithDensity}).
In this analysis, degeneracy between $B_\textmax$ and $\eta$ has been found: a larger $B_\textmax$ implies a larger $\eta$ and \textit{vice versa}.
The best fit gave $B_\textmax = 4.7\mu$G, $\eta = 0.5$, while $B_\textmax = 3.9\mu$G, $\eta = 0.4$ and $B_\textmax = 5.4\mu$G, $\eta = 0.7$ are still within 1 $\sigma$.
To illustrate the dependence of our approach on $B_\textmax$ and $\eta$, we will perform our analysis with these three pairs of values.
We furthermore adopt the values $r_c = (291 \pm 17)$ kpc, $n_\textmax = (3.44 \pm 0.04)\cdot 10^3$ m$^{-3}$, and $\beta = 0.75 \pm 0.03$ from the same work.
\\
In contrast, the Hydra A cluster is a cool-core cluster and exhibits magnetic field strengths of $6~\mu$G coherent on scales of 100 kpc and magnetic field strengths of 30 $\mu$G on scales of 4 kpc \cite{taylor:1993}.
The electron density in the cluster center is $n_\textmax \simeq 10^{4}$ m$^{-3}$ and the core radius is $r_c \simeq 130$ kpc \cite{davis:2009}.
We also adopt the value $\beta = 1$ from \cite{davis:2009}, while different values of $\eta$ have been used in the same work: mostly, $\eta = 0.9$ has been used, but also $\eta = 1/2, \eta = 2/3,$ and $\eta = 1$ have been considered.
Due to its strong influence on the possible limits on $g$, we will work with $\eta = 0.9$ as well as with $\eta = 2/3$.
The former value is suggested by observations of Abell 119 \cite{dolag:2001}, while the latter is predicted by flux conservation and is closer to the value observed in the Coma cluster.
\\
For both clusters, we will assume $y \simeq 10^{-5}$, a typical value of the Compton parameter observed in galaxy clusters \cite{planck2015SZ}.

%%%%%%%%%%%%%%%%%%%%%%%%%%%%%%%%%%%%%%%%%%%%%%%%%%%%%%%%%%%%%%%%%%%%%%%%%%%%%%%%%%%%%%%%%%%%%%%%%%%%%%%%%%%
\section{Results}

\subsection{Coupling constant constraints}
%%%%%%%%%%%%%%%%%%%%%%%%%%%%%%%%%%%%%%%%%%%%%%%%%%%%%%%%%%%%%%%%%%%%%%%%%%%%%%%%%%%%%%%%
%%%%%%%%%%%%%%%%%%%%%%%%%%%%%%%     Figure     %%%%%%%%%%%%%%%%%%%%%%%%%%%%%%%%%%%%%%%%%
%%%%%%%%%%%%%%%%%%%%%%%%%%%%%%%%%%%%%%%%%%%%%%%%%%%%%%%%%%%%%%%%%%%%%%%%%%%%%%%%%%%%%%%%
\begin{figure}[t]
\centering
{
\includegraphics[width=15cm]{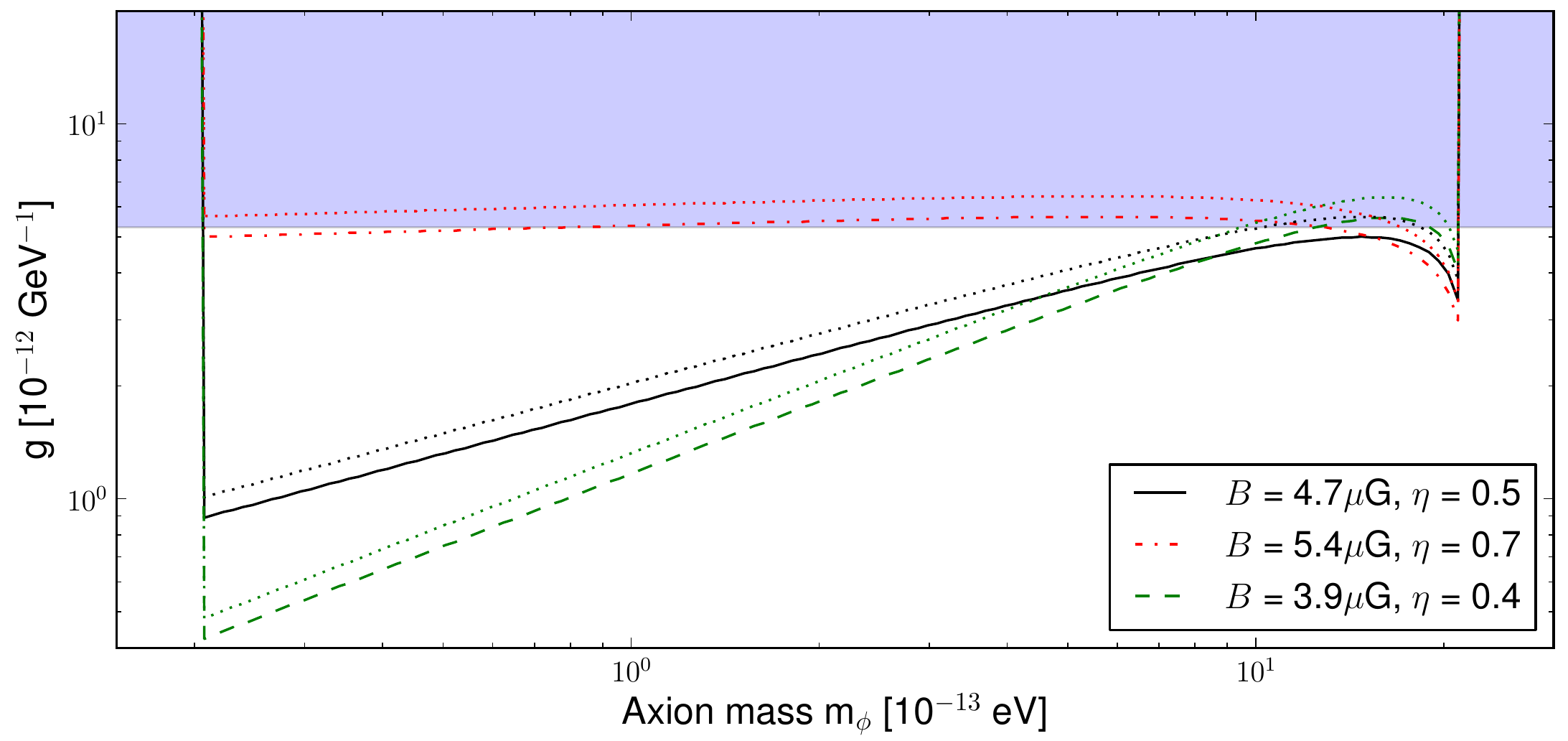}
}
\caption{
Obtained limits for different pairs of values for $B_\textmax, \eta$ obtained by the $\chi^2$-test with the data from the Coma cluster.
The black solid line shows the limits at 95\% C.L. when assuming the best-fit model from \cite{bonafede:2010}.
Using the upper and lower limits on $B_\textmax, \eta$ at $1\sigma$, one obtains the red dashed-dotted line and the green dashed line, respectively.
The limits at 99\% C.L. for the different values are indicated by the dotted lines.
While the limits on the coupling constant $g$ for high ALP masses (corresponding to resonances near the center of the cluster) are quite similar, the difference increases for smaller ALP masses (corresponding to resonances in the outer regions of the cluster).
The shaded area is excluded by limits obtained from SN1987A.}
\label{fig:etaComparison}
\end{figure}
%%%%%%%%%%%%%%%%%%%%%%%%%%%%%%%%%%%%%%%%%%%%%%%%%%%%%%%%%%%%%%%%%%%%%%%%%%%%%%%%%%%%%%%%
In figure (\ref{fig:etaComparison}), the limits obtained from the $\chi^2$-analysis of the Coma cluster are presented.
Due to the strong influence of $\eta$ on the magnetic field strength in the outer regions of the cluster, the limits from three different pairs of values for ($B_\textmax$, $\eta$) are shown.
These three pairs of values are the ones already presented in the previous section: (4.7$\mu$G, 0.5) provides the best fit to the observed rotation measures, while (3.9$\mu$G, 0.4) and (5.4$\mu$G, 0.7) are the lower and upper limits, respectively, at $1 \sigma$ confidence level.
\\
For the best-fit model from \cite{bonafede:2010}, the obtained limits are up to a factor of 5 stronger than the limits derived from SN1987A.
For $B_\textmax = 3.9\mu$G, $\eta = 0.4$, the obtained limits are even stronger for small ALP masses, while $B_\textmax = 4.7 \mu$G, $\eta = 0.7$ produces limits slightly weaker than the ones from SN1987A.
\\
In this context, a warning is necessary: the $\beta$-model of the free electron density as used in \cite{bonafede:2010} and adapted here is based on measurements of the Coma cluster's X-ray emission \cite{briel:1992}.
In this work, the largest distance from the cluster center probed is $\sim$15 $r_c$, as the signal becomes undetectable in the noise for larger distances.
This maximal tested radius corresponds to $n_e \simeq 8$ m$^{-3}$ and $m_\phi \simeq 1.0 \cdot 10^{-13}$ eV.
Although the density obviously has to decrease to the average cosmological density, the exact profile is not known.
We will assume that the $\beta$-profile holds down to the average cosmological density, but one should keep in mind that for $m_\phi \lesssim 10^{-13}$ eV, the limits are obtained under this assumption.
%%%%%%%%%%%%%%%%%%%%%%%%%%%%%%%%%%%%%%%%%%%%%%%%%%%%%%%%%%%%%%%%%%%%%%%%%%%%%%%%%%%%%%%%
%%%%%%%%%%%%%%%%%%%%%%%%%%%%%%%     Figure     %%%%%%%%%%%%%%%%%%%%%%%%%%%%%%%%%%%%%%%%%
%%%%%%%%%%%%%%%%%%%%%%%%%%%%%%%%%%%%%%%%%%%%%%%%%%%%%%%%%%%%%%%%%%%%%%%%%%%%%%%%%%%%%%%%
\begin{figure}[t]
\centering
{
\includegraphics[width=15cm]{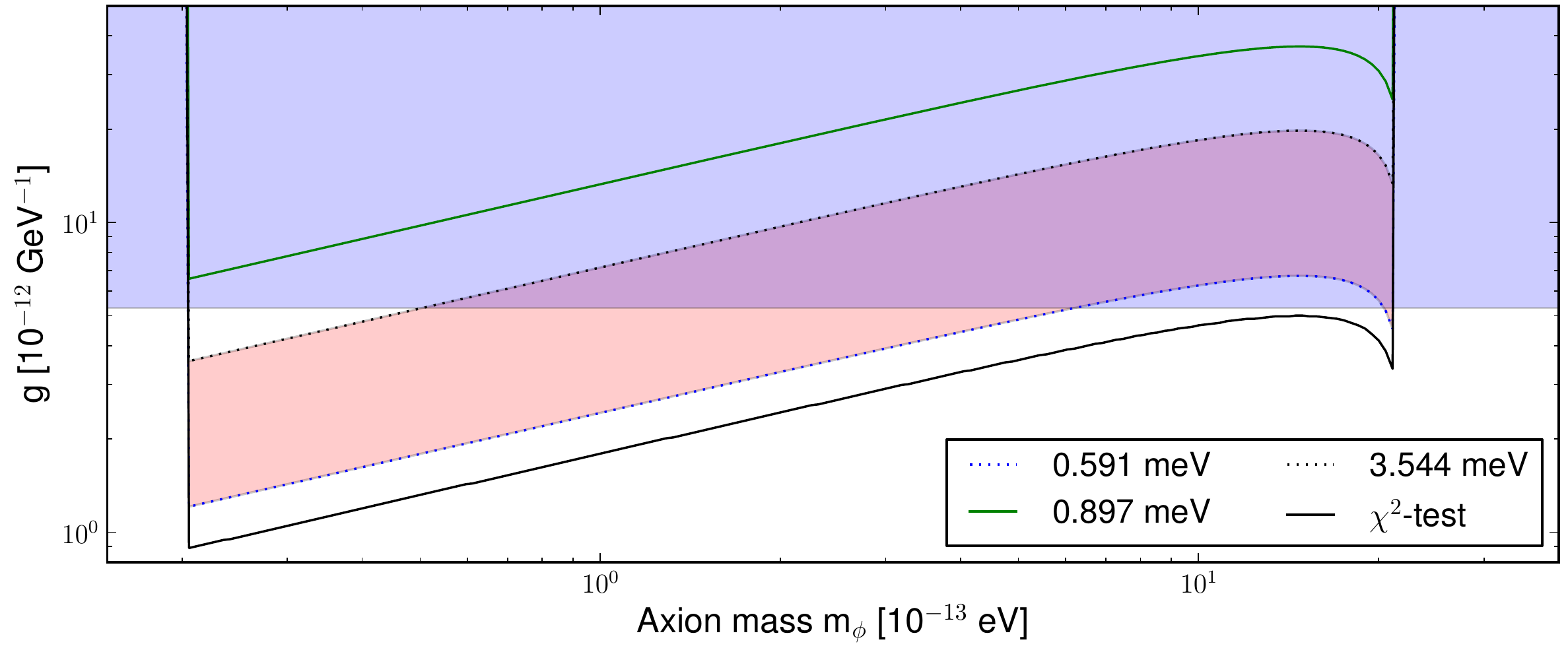}
}
\caption{
Comparison of the limits at 95\% C.L. obtained by the $\chi^2$-test (\ref{formula:chiSqTest}) and the direct temperature comparison (\ref{formula:temperatureComparison}) with different photon energies.
For the photon energy $\omega$ = 8.97$\cdot 10^{-4}$ eV (green solid line), the function $h(\omega)$ in eq. (\ref{formula:h}) becomes almost zero. But the rapid change of the function causes a high uncertainty, effectively even weakening the obtained limits.
For all other (center-) frequencies measured by \textit{Planck}, the obtained limits lie inside the red/light gray area, where $\omega = 35.44\cdot 10^{-4}$ eV gives the weakest bounds and $\omega = 5.91 \cdot 10^{-4}$ eV gives the strongest bounds.
These bounds are approximately 20\% weaker than the limits obtained by the $\chi^2$-test (black solid line), proving that the direct temperature comparison provides realistic results.
The blue/dark gray shaded area is excluded by limits obtained from SN1987A.
For this plot, the data of the Coma cluster and $\eta$ = 1/2 has been used.
}
\label{fig:gBoundsDifferentOmega}
\end{figure}
%%%%%%%%%%%%%%%%%%%%%%%%%%%%%%%%%%%%%%%%%%%%%%%%%%%%%%%%%%%%%%%%%%%%%%%%%%%%%%%%%%%%%%%%
\\
In figure (\ref{fig:gBoundsDifferentOmega}), a comparison of the limits obtained by the $\chi^2$-test (\ref{formula:chiSqTest}) and the limits obtained by the temperature comparison (\ref{formula:temperatureComparison}) is presented.
For the limits by the temperature comparison, the value of $g$ was determined according to eq. (\ref{eq:finalConvProb}).
Additionally, we included Gaussian error propagation to estimate the uncertainty and to obtain limits at different confidence levels.
We adopted the uncertainties given by \cite{bonafede:2010}, i.e. $\sigma_{B_\textmax}/B_\textmax = 16\%$, $\sigma_{n_\textmax}/n_\textmax = 1.1\%$, $\sigma_{\beta}/\beta = 4\%$, $\sigma_{r_c}/r_c = 6\%$, as well as $\sigma_\omega/\omega = 20\%$(low frequency instrument)/33\%(high frequency instrument) from \cite{planck2010HFI}.
Although the total conversion probability (\ref{eq:finalConvProb}) does not dependent explicitly on $N$ anymore, the multiple resonances induce an additional uncertainty.
When assuming $N \gtrsim 40$ (see above), and $\sigma_N \propto \sqrt{N}$, one obtains $\sigma_N/N \simeq 16\%$.
As $\sum_i R_i \propto N \propto R_\beta \propto r_c$, we absorb this uncertainty into $\sigma_{r_c}/r_c$ and conservatively set $\sigma_{r_c}/r_c \simeq 25\%$.
Finally, we conservatively set $\sigma_y/y = 50\%$.
The error is usually dominated by the uncertainty of the photon energy $\omega$; only for very low and for very high photon energies, the uncertainty is dominated by $\sigma_y$.
\\
In order to avoid the singular behavior of $R_\beta$ for $n_\res \rightarrow n_c$, we exclude the innermost region with $n_\res / n_c \gtrsim 10\%$, and, due to its large influence, we fix $\eta = 1/2$.
This plot should therefore demonstrate the robustness of the obtained limits with respect to astrophysical uncertainties.
\\
The used photon energy determines the numerical value of the function $h(\omega)$, see eq. (\ref{formula:h}). This function is usually of order unity, but reaches zero for $\omega = 3.83\,T_{\text{CMB}}$.
Naively, one could expect to obtain arbitrarily strong limits when simply using a photon energy close to this value. 
This is, however, unphysical: 
the \textit{Planck} high-frequency channels have bandwidths of $\Delta \nu / \nu \approx 0.33$ \cite{planck2010HFI}, meaning that every frequency map is actually an average over a range of frequencies.
One therefore also would have to take an appropriate average over the function $h(\omega)$, preventing arbitrarily small limits.
%%%%%%%%%%%%%%%%%%%%%%%%%%%%%%%%%%%%%%%%%%%%%%%%%%%%%%%%%%%%%%%%%%%%%%%%%%%%%%%%%%%%%%%%
%%%%%%%%%%%%%%%%%%%%%%%%%%%%%%%     Figure     %%%%%%%%%%%%%%%%%%%%%%%%%%%%%%%%%%%%%%%%%
%%%%%%%%%%%%%%%%%%%%%%%%%%%%%%%%%%%%%%%%%%%%%%%%%%%%%%%%%%%%%%%%%%%%%%%%%%%%%%%%%%%%%%%%
\begin{figure}[ht]
\centering
{
\includegraphics[width=15cm]{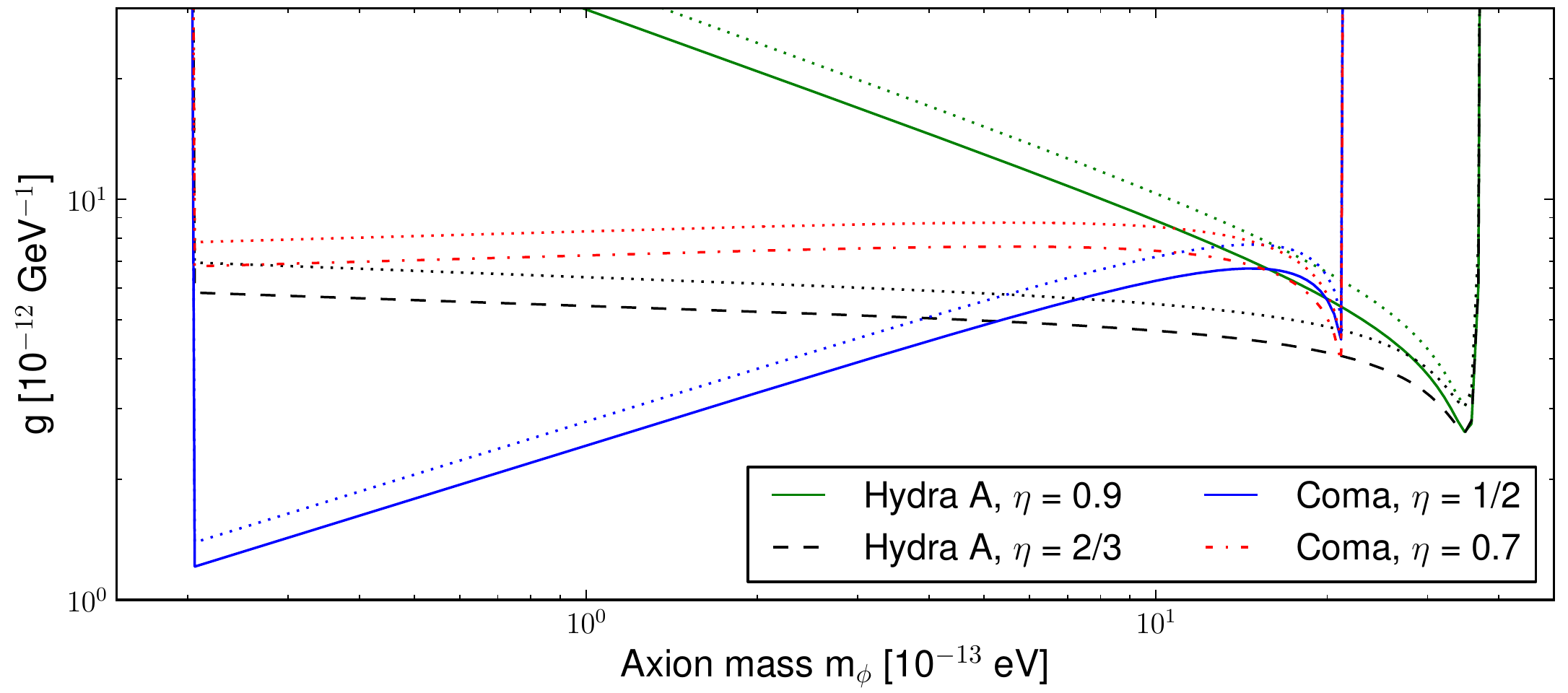}
}
\caption{
Comparison of the obtained limits at 95\% C.L. for two different clusters: the Coma cluster and the Hydra A cluster.
Additionally to each labeled line, the limits at 99\% C.L. are shown as dotted lines. 
The higher central density of the Hydra A cluster allows limits for slightly higher masses, while the strength of the limits strongly depends on the assumed values of $\eta$.
For this plot, $\omega = 5.9\cdot 10^{-4}$ eV has been used.
}
\label{fig:ComaHydra}
\end{figure}
%%%%%%%%%%%%%%%%%%%%%%%%%%%%%%%%%%%%%%%%%%%%%%%%%%%%%%%%%%%%%%%%%%%%%%%%%%%%%%%%%%%%%%%% 
\\
A comparison of the limits obtained from the Coma cluster and from the Hydra A cluster using eq. (\ref{eq:finalConvProb}) is shown in figure (\ref{fig:ComaHydra}).
For the Hydra A cluster, we have assumed 25\% uncertainties for all cluster parameters, i.e. $r_c$, $n_\textmax$, $B_\textmax$, $\beta$, and kept $\sigma_y = 50\%$ as before.
The higher central electron density in the Hydra A cluster allows higher ALP masses to undergo resonant conversion, therefore slightly expanding the mass range accessible for the method presented here.
The higher value of $\eta$ leads to a faster decrease of the magnetic field strength with increasing radius, such that the obtained limits become weaker for smaller ALP masses.
For illustration, we also show the limits for the Hydra A cluster with $\eta = 2/3$ and the Coma cluster with $\eta = 0.7$: in this case, the radial decrease of the magnetic field is very similar, while the higher value of the central magnetic field in the Hydra A cluster leads to a higher conversion probability and therefore slightly stronger limits.

%%%%%%%%%%%%%%%%%%%%%%%%%%%%%%%%%%%%%%%%%%%%%%%%%%%%%%%%%%%%%%%%%%%%%%%%%%%%%%%%%%%%%%%% 
\subsection{Future perspective}
\begin{figure}[t]
\centering
{
\includegraphics[width=15cm]{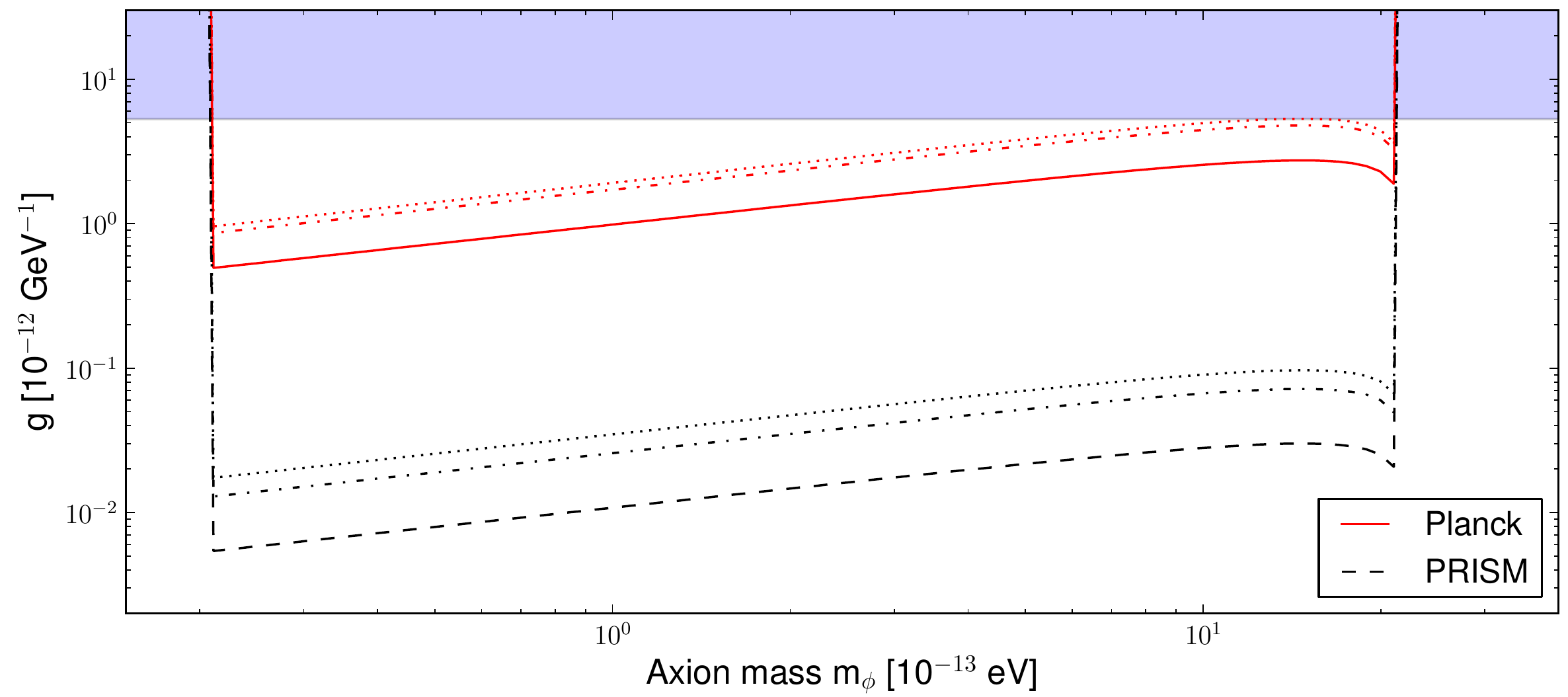}
}
\caption{
Projected sensitivity of the presented approach when using the full \textit{Planck} data or data from a future, PRISM-like experiment.
The sensitivity of \textit{Planck} with no error penalty is shown with the solid red line, while the dashed black line corresponds to a PRISM-like experiment.
The dashed-dotted lines are obtained when an error penalty of 5 is used, the dotted lines for error penalties of 10.
The shaded area is excluded by limits obtained from SN1987A.
For this plot, the parameters of the Coma cluster have been used.
}
\label{fig:Future}
\end{figure}

In a more detailed study, one would have to simultaneously fit several contributions to the recorded data, e.g. the tSZ effect, thermal dust and synchrotron radiation.
Including photon-ALP conversion in this procedure, one would then obtain limits on the coupling constant $g$.
To estimate the possible limits with this approach, we restrict ourselves to a simpler approach:
we simulate a tSZ-signal according to eq. (\ref{formula:tSZ}), where we use the uncertainties of the \textit{Planck}-experiment, multiplied with a factor referred to as ``error penalty''.
This error penalty parametrizes the additional uncertainty induced by subtracting the foreground emission.
In a second step, we perform a $\chi^2$-analysis, according to eq. (\ref{formula:chiSqTest}), where we fit both a tSZ-signal and photon-ALP conversion to the simulated signal.
We thus obtain an upper limit on the sensitivity for the coupling constant $g$.
\\
We use the temperature sensitivities described in \cite{planck2010LFI} and \cite{planck2010HFI}, where the sensitivities range between $\Delta T/T \simeq 2.2\cdot 10^{-6}$ for $\nu = 143$ GHz and $\Delta T/T \simeq 6 \cdot 10^{-3}$ for $\nu = 857$ GHz.
In figure (\ref{fig:Future}), we show the possible limits with error penalties of 1, 5, and 10, where the parameters of the Coma cluster have been used and we averaged the limits obtained from ten different simulated realizations.
\\
One also can extent this approach to proposed future experiments for highly sensitive CMB observation like PRISM \cite{PRISM} or PIXIE \cite{PIXIE}.
PRISM is proposed to have 32 broad-range frequency channels as well as 300 narrow frequency channels covering the range from 30 GHz to 6000 GHz.
The simulated 4-year sensitivities range from
$\delta I_\nu$ = 3.6$\times10^{-27}$ Wm$^{-2}$Hz$^{-1}$sr$^{-1}$ for frequencies between 30 GHz and 180 GHz
to $\delta I_\nu$ = 1.6$\times10^{-26}$ Wm$^{-2}$Hz$^{-1}$sr$^{-1}$ for frequencies greater than 3 THz.
Performing the same analysis as before, we arrive at sensitivities of $\gtrsim 10^{-14}$GeV$^{-1}$, two orders of magnitude below the currently strongest limits for this mass range.
This result is also displayed in figure (\ref{fig:Future}).
PIXIE is proposed to have 400 channels covering the same frequency range as PRISM, reaching slightly worse sensitivities than PRISM.
We therefore expect PIXIE to be sensitive to values of $g$ very similar to the ones presented for PRISM.

%%%%%%%%%%%%%%%%%%%%%%%%%%%%%%%%%%%%%%%%%%%%%%%%%%%%%%%%%%%%%%%%%%%%%%%%%%%%%%%%%%%%%%%%
%%%%%%%%%%%%%%%%%%%%%%%%%%%%%%%     Figure     %%%%%%%%%%%%%%%%%%%%%%%%%%%%%%%%%%%%%%%%%
%%%%%%%%%%%%%%%%%%%%%%%%%%%%%%%%%%%%%%%%%%%%%%%%%%%%%%%%%%%%%%%%%%%%%%%%%%%%%%%%%%%%%%%%
%%%%%%%%%%%%%%%%%%%%%%%%%%%%%%%%%%%%%%%%%%%%%%%%%%%%%%%%%%%%%%%%%%%%%%%%%%%%%%%%%%%%%%%% 

%%%%%%%%%%%%%%%%%%%%%%%%%%%%%%%%%%%%%%%%%%%%%%%%%%%%%%%%%%%%%%%%%%%%%%%%%%%%%%%%%%%%%%%%%%%%%%%%%%%%%%%%%%%
\section{Discussion and Conclusions}

In this study, we have shown that \textit{Planck's} recent measurements of the tSZ Compton-parameter $y$ \cite{planck2015SZ} can be used to constrain the coupling constant $g$ of pseudoscalar ALPs to photons.
To this end, we compared the temperature change due to the tSZ-effect of the CMB photons reaching us from galaxy clusters with the temperature change due to resonant photon-ALP conversion.
Photon-ALP conversion is most effective at resonance where it leads to the strongest limits on the coupling constant.
On the other hand, resonant photon-ALP conversion is only possible for a limited range of ALP masses, typically of the order of 10$^{-13}$ eV; this range depends on the effective photon mass in the galaxy clusters, and therefore on the free electron density.
\\
The strength of the obtained limits depends both on the density profile in the galaxy cluster as well as on the magnetic field.
In our study, we used a $\beta$-model, extended with density modulations, for these profiles, and typical values for the magnetic field strength, electron densities and the observed $y$-parameter for galaxy clusters.
Under these assumptions, we can derive limits on the photon-ALP coupling constant $g$, which are slightly stronger than the existing bounds in this mass region from SN1987A \cite{sn1987:3}, and furthermore provide an independent constraint.
\\
The scaling of the magnetic field with the electron density has a strong influence on the limits for smaller ALP masses.
Further investigations of the magnetic field strength in the outer regions of galaxy clusters or the selection of suitable clusters will be necessary to solve this problem and might provide stronger limits.
Another approach might be to consider the magnetized jets emitted by AGNs.
\\
We estimated the parameter region a PRISM-like experiment would be sensitive to and found that for the mass range given by the resonance condition, sensitivities down to $g \gtrsim 10^{-14}$ GeV$^{-1}$ are realistic.
This is especially interesting, as this is part of the parameter space has been invoked to explain the soft X-ray excess of the Coma cluster \cite{Angus:2014} or the anomalous gamma-ray transparency \cite{Meyer:2013}.

\section*{Acknowledgements}
We would like to thank Anne-Christine Davis, Alexandre Payez and Andreas Ringwald for useful comments.
This work was supported by the Deutsche Forschungsgemeinschaft (DFG) through the Collaborative Research Centre SFB 676 ``Particles, Strings and the Early Universe''. 
Furthermore, we acknowledge support from the Helmholtz Alliance for Astroparticle Physics (HAP) funded by the Initiative and Networking Fund of the Helmholtz Association.

%%%%%%%%%%%%%%%%%%%%%%%%%%%%%%%%%%%%%%%%%%%%%%%%%%%%%%%%%%%%%%%%%%%%%%%%%%%%%%%%%%%%%%%%%%%%%%%%%%%%%%%%%%
%%%%%%%%%%%%%%%%%%%%%%%%%%%%%%%%%%%%%%%%     Appendix      %%%%%%%%%%%%%%%%%%%%%%%%%%%%%%%%%%%%%%%%%%%%%%%
%%%%%%%%%%%%%%%%%%%%%%%%%%%%%%%%%%%%%%%%%%%%%%%%%%%%%%%%%%%%%%%%%%%%%%%%%%%%%%%%%%%%%%%%%%%%%%%%%%%%%%%%%%
\appendix
\section{Conversion probability in case of multiple level crossings}
In \cite{parke:1986} it was derived that the photon-ALP conversion probability is given by
\begin{equation}
\label{eq:ConversionProb}
P_{\gamma \rightarrow \phi} = \frac{1}{2} + \left( p - \frac{1}{2} \right) \cos ( 2 \tilde{\theta}_0 ) 
\cos ( 2 \tilde{\theta}_D ),
\end{equation}
where $\tilde{\theta}_0$ is the effective mixing angle at the production of the CMB, $\tilde{\theta}_D$ is the effective mixing angle at the detection, and $p$ is the total level crossing probability.
When averaging over the distances between the individual resonances, the total level crossing probability 
is the sum of all the individual probabilities of getting an odd number of level crossings, so simply the classical probability results.
\\
We will first consider an odd number of resonances.
In the case of only one resonance, the level crossing probability is just given by this single level crossing probability.
In case of three level crossings, one has
\begin{equation}
\label{eq:LevelCrossing3}
p = p_1 \left[ p_2 p_3 + (1-p_2) (1-p_3) \right] + (1-p_1) \left[ p_2 (1-p_3) + (1-p_2) p_3 \right],
\end{equation}
where $p_i$, $i$ = 1, 2, 3, are the level crossing probabilities for the individual resonances.
These individual probabilities can be calculated with the Landau-Zener formula \cite{parke:1986}
\begin{equation}
\label{eq:LandauZenerAppendix}
p_i \simeq \exp(- 2 \pi R k \sin^2 \theta_{\text{res}} ) =: \exp( - \delta_i),
\end{equation}
where $R$ is the scale parameter defined in eq. (\ref{formula:scaleParameter}), $k$ is the wavenumber of the photon-ALP conversion (\ref{eq:k}) and $\theta_{\text{res}}$ is the magnetic mixing angle given in eq. (\ref{eq:thetaNoPlasma}), evaluated at resonance.
Note that this formula only holds when the electron density varies roughly linearly within the resonance.
In our study, the exponent $\delta_i$ is very small, so one can approximate $\exp(-\delta_i) \simeq 1- \delta_i$. Plugging this into formula (\ref{eq:LevelCrossing3}), one obtains
\begin{equation}
p = 1 - \sum_{i=1}^{3} {\delta_i} + \sum_{i \neq j} \delta_i \delta_j - 4 \cdot \delta_1 \delta_2 \delta_3.
\end{equation}
This result generalizes to any odd number $n$ of resonances:
\begin{equation}
\label{eq:pOdd}
p = 1 - \sum_{i=1}^{n} {\delta_i} + \mathcal{O}(\delta^2 + \text{higher}).
\end{equation}
In case of an odd number of resonances, one necessarily has $\cos(2 \tilde{\theta}_{i}) < 0$ and $\cos(2 \tilde{\theta}_{i}) > 0$ or
\textit{vice versa}.
Parametrizing $\cos(2 \tilde{\theta}_{0}) = \pm (-1 + \epsilon_0)$, $\cos(2 \tilde{\theta}_{D})= \pm ( 1 - \epsilon_D)$ with $|\epsilon_0|, |\epsilon_D| \leq 1$, one has 
\begin{equation}
\cos ( 2 \tilde{\theta}_0 ) \cos ( 2 \tilde{\theta}_D ) = -1 + \epsilon_0 + \epsilon_D - \epsilon_0 \epsilon_D.
\end{equation}
Using the last two formulas, the conversion probability (\ref{eq:ConversionProb}) for an odd number of resonances reads 
\begin{equation}
P_{\gamma \rightarrow \phi} = \sum_{i}{\delta_i} + (\epsilon_0 + \epsilon_D) \left( 1/2 - \sum_{i}{\delta_i} \right)
- \epsilon_0 \epsilon_D \left( 1/2 - \sum_{i}{\delta_i} \right) + \mathcal{O}(\delta^2 + \text{higher}).
\end{equation}
In our study, we have four resonances. In this case, the total level crossing probability becomes
\begin{equation}
\begin{split}
p = p_1 \{ p_2 \left[ p_3 (1-p_4) + (1-p_3) p_4 \right] +
 (1-p_2) \left[  p_3 p_4 + (1-p_3) (1-p_4) \right] \} + \\
+ (1-p_1) \{ p_2 \left[  p_3 p_4 + (1-p_3) (1- p_4) \right] + 
(1-p_2) \left[ p_3 (1-p_4) + (1-p_3)p_4 \right] \},
\end{split}
\end{equation}
where $p_i$ are again the individual level crossing probabilities.
Using formula (\ref{eq:LandauZenerAppendix}) and $\delta_i \ll 1$ as before, one arrives at
\begin{equation}
p = \sum_{i=1}^{4} {\delta_i} -  \sum_{i \neq j}{\delta_i \delta_j} + 4 \cdot \sum_{i = 1}^{4}{(\delta_1 \delta_2 \delta_3 \delta_4)/\delta_i}
- 8 \cdot \delta_1 \delta_2 \delta_3 \delta_4.
\end{equation}
More general, for any even number $n$ of resonances, the total level crossing probability will be
\begin{equation}
p = \sum_{i=1}^{n} {\delta_i} + \mathcal{O}(\delta^2 + \text{higher}).
\end{equation}
In case of an even number of resonances, one either starts below the resonance density, and also ends below the resonance density, or one starts and ends above the resonance density.
One can therefore parametrize $\cos(2 \tilde{\theta}_{0}) = \pm (-1 + \epsilon_0)$, $\cos(2 \tilde{\theta}_{D})= \pm (-1 + \epsilon_D)$ with $|\epsilon_0|, |\epsilon_D| \leq 1$, and obtains 
\begin{equation}
\cos ( 2 \tilde{\theta}_0 ) \cos ( 2 \tilde{\theta}_D ) = 1 - \epsilon_0 - \epsilon_D + \epsilon_0 \epsilon_D.
\end{equation}
Using the last two formulas, conversion probability becomes
\begin{equation}
P_{\gamma \rightarrow \phi} = \sum_{i}{\delta_i} + (\epsilon_0 + \epsilon_D) \left( 1/2 - \sum_{i}{\delta_i} \right)
- \epsilon_0 \epsilon_D \left( 1/2 - \sum_{i}{\delta_i} \right) + \mathcal{O}(\delta^2 + \text{higher}),
\end{equation}
which is, up to terms of $\mathcal{O}(\delta^2$ + \text{higher}), the same formula as before.
\\
As it was argued in \cite{resPhotonAxion}, due to the high electron density at the time of recombination $\cos( 2 \tilde{\theta}_0) = -1 + \epsilon_0$ is very close to $-1$, with
\begin{equation}
\epsilon_0 \simeq 7 \cdot  10^{-18} \cdot \left( \frac{g}{10^{-13} \text{ GeV}^{-1}} \frac{B_{\text{cosmological}}}{\text{nG}} \frac{\omega}{T_{\text{CMB}}} \right) ^2
\end{equation}
where $B_{\text{cosmological}}$ is the magnetic field strength on cosmological scales and all values are taken today.
Using a free electron density of about 1 cm$^{-3}$ = 10$^6$ m$^{-3}$ inside our galaxy, one obtains $\cos( 2 \tilde{\theta}_D) = -1 + \epsilon_D$ with
\begin{equation}
\epsilon_D \simeq 5\cdot 10^{-15} \cdot \left( \frac{m_{\phi}}{10^{-14} \text{ eV}} \right)^4 + 7 \cdot  10^{-12} \cdot \left( \frac{g}{10^{-13} \text{ GeV}^{-1}} \frac{B_{\text{galactic}}}{\mu \text{G}} \frac{\omega}{T_{\text{CMB}}} \right)^2.
\end{equation}
The first term can reach values up to $10^{-5}$ for the ALP-masses considered here.
Therefore, all terms of order $\epsilon^2$ and $\delta \cdot \epsilon$ in the conversion probability can certainly be neglected and, as $\delta_i \ll 1$, also all terms of order $\delta^2$.
The remaining result simply is
\begin{equation}
P_{\gamma \rightarrow \phi} \simeq \sum_{i}{\delta_i} + (\epsilon_0 + \epsilon_D)/2.
\end{equation}
In this study, we will always make the conservative assumption that 
\begin{equation}
P_{\gamma \rightarrow \phi} \simeq \sum{\delta_i},
\end{equation}
therefore conservatively underestimating the conversion probability,
since any possible (positive) contribution from $\epsilon_{0}$ and $\epsilon_{D}$ is neglected.
As we are interested in upper limits for $g$, underestimating $P_{\gamma \rightarrow \phi}$ is conservative and only makes the limits more reliable.

%%%%%%%%%%%%%%%%%%%%%%%%%%%%%%%%%%%%%%%%%%%%%%%%%%%%%%%%%%%%%%%%%%%%%%%%%%%%%%%%%%%%%%%%%%%%%%%%%%%%%%%%%%
%%%%%%%%%%%%%%%%%%%%%%%%%%%%%%%%%%%%% References %%%%%%%%%%%%%%%%%%%%%%%%%%%%%%%%%%%%%%%%%%%%%%%%%%%%%%%%%
%%%%%%%%%%%%%%%%%%%%%%%%%%%%%%%%%%%%%%%%%%%%%%%%%%%%%%%%%%%%%%%%%%%%%%%%%%%%%%%%%%%%%%%%%%%%%%%%%%%%%%%%%%

\end{document}